\documentclass[onecolumn,12pt]{article}
\title{Phase behavior under the averaging over disorder realizations}

\author{E.~V.~Vakarin$^a$, W.~Dong$^b$, J.~P.~Badiali$^a$\\
$^a$ UMR 7575 LECIME, ENSCP-UPMC-CNRS, 11\\
\vspace{0.3cm} 
rue P. et M. Curie, 75231 Cedex 05, Paris, France\\
$^b$ Laboratoire de Chimie, UMR 5182 CNRS,\\ 
Ecole Normale
Sup\'{e}rieure de Lyon,\\ 46 All\'{e}e
d'Italie, 69364 Lyon, Cedex 07, France}
\date{}

\begin{document}


\maketitle

\begin{abstract}
Effects of the averaging over disorder realizations (samples) on the phase
behavior are analyzed in terms of the mean field approximation for the 
random field Ising model with infinite range interactions. 
It is found that the averaging is equivalent to a 
drastic modification in the statistics of the quenched variables. In its 
turn, this lowers the critical temperature of a second-order phase 
transition or, depending on the sampling, even suppresses the ordered 
phase. Possible first order transitions  are shown to be softened by the sample 
averaging. Common issues and differences in the interpretation of these 
effects in the context of the simulation and experimental studies are 
discussed. 
\end{abstract}

\vspace{1cm}

\vspace{0.5cm}

\section{Introduction}
An influence of a quenched disorder on the phase behavior 
is a long-standing problem that has received much interest\cite{RF,SG}.
Various theoretical aspects, such as disorder relevance,
Griffiths phenomena, possible lack of self-averaging\cite{Ahaself} or 
disorder-induced rounding are being discussed\cite{Vojta} 
(see \cite{RVojta} for a recent review). In the case of the so-called
weak (or random $T_c$) disorder it is believed\cite{RVojta} that the
details of the probability distribution should not be important because the
physics is dominated by the long-range properties.

On the other hand, there are systems in which the shape of the probability 
distribution is important. For instance, the fluid 
adsorption\cite{hysteresis} or the colloid-polymer mixtures separation in 
random porous media are found\cite{BL} to belong to the universality class 
of the random-field Ising model (RFIM)\cite{RF,RFIM}. The latter is just a 
prototype of a system, where the randomness plays an essential role, the 
so-called strong disorder\cite{Igloi}. In contrast to the weak 
disorder case\cite{Vojta,RVojta}, a change in the distribution modifies the 
degree of frustration, inducing specific fluctuations\cite{ParSour},  which 
are not present in the pure (bulk) case. The RFIM has been analyzed in the 
mean field approximation (MFA) \cite{Pytte,Maritan,Aharony} and more 
recently by computer simulations\cite{ParSour,WM}. The distribution 
width\cite{Pytte}, asymmetry\cite{Maritan} or bimodality\cite{Aharony,ceva,fytas} have 
been shown to induce crucial changes in the phase diagram. 
Nevertheless, in the context of adsorption studies it becomes questionable
\cite{hysteresis,avalanches} that the true equilibrium picture is observed
in experiments, revealing a hysteretic behavior instead of sharp steps
typical for a two phase coexistence. Moreover, it has been demonstrated
\cite{hysteresis,Kierlik} that the hysteresis can exist even without an
underlying phase transition. This, however, does not discard completely a 
possibility of the phase behavior in disordered systems, but rather makes 
one to search for the conditions at which it can be observed.    

Another aspect of the randomness is that two quenched systems prepared in 
the same way are not microscopically identical because only few macroscopic
disorder parameters are usually controlled. Therefore, one has to average
over different realizations.   
In the present study we argue that in this class of systems the 
averaging over disorder realizations in the 
simulations or appearing in the context of self-averaging 
hypothesis\cite{RF,SG} in the analysis of experiments, is equivalent to 
crucial changes in the distribution of quenched variables (random fields in 
the RFIM language). In this way a sample averaged distribution brings up new 
statistical features which are absent from the statistics of each sample. 
Consequently, the phase behavior seen on average can be remarkably different 
from that in a separate sample. In particular, we are going to analyze 
whether a small difference of realizations can lead to qualitative 
changes in the phase behavior.  
 
\section{Averaging in disordered systems}
Let us consider a quenched-annealed system. The annealed 
subsystem consists of interacting species (e. g. particles or spins) 
characterized by a set of variables $\{s_i\}$. The latter are governed by 
some Hamiltonian $H[\{s_i\}|\{h_k\}]$ that contains a set of quenched 
variables (local porosities or magnetic fields) $\{h_k\}$, distributed 
according to some probability density $f(\{h_k\})$. For simplicity we 
assume the disorder to be non-correlated $f(\{h_k\})=\prod_k f(h_k)$, such
that the position label is dropped $f(h_k)=f(h)$. In contrast to coupled 
annealed systems (e.g. an adsorbate-substrate coupling\cite{PRL}), in 
quenched-annealed systems the averaging should be taken in two steps.   
Tracing off the annealed variables one gets a quenched
thermodynamic function $m(h)$  (an order parameter) that is conditional to 
the state, $h$, of the quenched counterpart. Then, as usually\cite{RF,SG}, 
one takes an average of $m(h)$ with the distribution $f(h)$.

However, in reality, the situation is not so straightforward. In experimental
or simulation studies only few disorder parameters (e. g. mean porosity, 
average field or site activity) are usually controlled during the quenching.
Therefore, one has to deal with sample-to-sample fluctuations in
different realizations $r$. Thus, for a given sample the distribution
implicitly depends on the realization $f(h)=f(h|r)$. Consequently, the 
disorder averages are also realization dependent 
\begin{equation}
m(r)={\overline{m(h)}}=\int dh f(h|r) m(h)
\end{equation} 
In order to find a "representative" result in the simulations one could 
generate an ensemble of samples corresponding to different realizations 
which appear with a probability density $\varphi(r)$. Such a procedure
has recently been reported\cite{simsample}.  
Then 
an average over realizations is taken  
\begin{equation}
m=\int dr\varphi(r)m(r)=\int dr\varphi(r)\int dh f(h|r)m(h)
\end{equation}
If the sampling $\varphi(r)$ does not depend on the $h$-disorder,
we may change the order of integration, defining the realization-averaged
(or sample-averaged) distribution 
\begin{equation}
\label{reals}
\Psi(h)=\int dr \varphi(r)f(h|r).
\end{equation}
Then the order parameter should be calculated from
\begin{equation}
\label{obs}
m=\int dh \Psi(h) m(h).
\end{equation}
On the other hand, an experimentalist is usually working with a single
system (or with few systems). This rises a question on a representability
of a single experiment  for a given class of materials. This leads to the
concept of self-averaging\cite{RF,SG,Ahaself,RVojta}. Namely, it is assumed
that an experimental system is large enough, such that it may be viewed
as a composition of a large number of macroscopic subsystems (samples). Each 
of them corresponds to a different realization of the quenched disorder. In 
this way a measurement corresponds to an average over the subsystems, thus we
arrive at the same representation (\ref{reals}), (\ref{obs}).
This mathematical similarity enables one to compare the simulation
and the experimental results. Note that the averaging over realizations
(\ref{reals}) is formally equivalent to creating some "representative"
or "typical"\cite{RF} sample with a distribution $\Psi(h)$ that is thought 
to incorporate the generic features, independent of the system preparation 
procedure. However, in the context of experimental studies, invoking the 
self-averaging concept, this system is (or is believed to be) a real object, 
while in the simulation, where each realization can be tested independently, 
it is artificial. 

Therefore, despite its logical transparency, the averaging  (\ref{reals})
is not a trivial procedure. It leaves some important questions, such as
how the final result (\ref{obs}) depends on the sampling $\varphi(r)$
and on the quenching in each sample $f(h|r)$? Indeed, from 
eq.~(\ref{reals}) we may expect that, depending on these ingredients, the 
resulting distribution $\Psi(h)$ could be remarkably different from that in 
a separate sample. From a quite general point of view this issue has been 
analyzed within the superstatistical approach\cite{Beck}. It has been 
demonstrated how an exponential distribution (e.g. of Boltzmann type) 
transforms into a power-law as a result of a fluctuating 
environment\cite{Beck,Wilk} or a constraint\cite{PRE-power} restricting the 
phase space. 
In its turn, such 
a modification in the statistics of the quenched variables 
(namely, an enhancement of "rare" events \cite{Vojta,RVojta,Igloi}) might 
lead to non-trivial consequences for the thermodynamics, especially near a 
phase transition point. In what follows we analyze this issue in more 
details.

\section{Phase behavior}
Realizations of the quenched subsystem are usually obtained by repeating
some preparation procedure (e.g. sol-gel technique in experiments, or 
diffusion-limited aggregation in the simulation studies). Therefore, it is 
reasonable to accept that in all samples the distribution $f(h|r)$ has the 
same functional form. For analytical purposes let it be gaussian  
\begin{equation}
\label{fg}
f(h|r)=\left[\sqrt{2\pi r^2}
\right]^{-1}
\exp{(-h^2/2r^2)}
\end{equation}
with zero mean ${\overline{h}}=0$ and dispersity $r$ that may vary
from sample to sample. As is discussed above, this variation is governed by 
the sampling $\varphi(r)$. One may argue that taking a gaussian distribution
with varying dispersity is not realistic because the latter can also be 
controlled and then all realizations would be identical. In this respect
it should be noted that real (e.g. pore size) distributions are usually more
complex than gaussian. Therefore, even if the first two moments are fixed, 
the other parameters (asymmetry, multimodality, etc.) are not controlled.
Thus, our choice (\ref{fg}) is just the simplest way of modeling such a 
situation.

\subsection{Discrete sampling}
One of the simplest non-trivial samplings is a
discrete bimodal $\varphi(r)=\varphi_d(r)$
\begin{equation}
\varphi_d(r)=c\delta(r-a)+(1-c)\delta(r-a-\Delta),
\end{equation}
where we have two sample populations, one with $r=a$ and
concentration $c$, the other with $r=a+\Delta$ and concentration
$1-c$. From eq.(\ref{reals}) we get
\begin{equation}
\label{pd}
\Psi_d(h)=cf(h|a)+(1-c)f(h|a+\Delta).
\end{equation}
This function is displayed in Fig.~1 in comparison with a gaussian 
distribution of the same width (i.e. ${\overline{h^2}}$ is the same). It is 
seen that $\Psi_d(h)$ has a pronounced non-gaussian shape with a central 
peak (around $h=0$) and the tail that favors higher $h$ values. In order to 
estimate the effect of this  distribution on the criticality we consider the 
RFIM \cite{RF,RFIM} with {\it infinite-range 
interactions} \cite{Pytte,Maritan,Aharony}. This is a generic toy model exhibiting 
a second-order phase transition between the ferromagnetic ($m\ne 0$) and 
"frozen" ($m=0$) or independent spin\cite{Pytte,Maritan,Aharony} phases. We 
are focusing on the very existence of the transition, without discussing its 
universality class (critical exponents, etc.). For this reason we deal the 
MFA which is sufficient for a qualitative estimation. In addition it is  
expected\cite{Pytte,Maritan} to be exact for the infinite-range 
interactions. Therefore, we have to solve 
\begin{equation}
\label{MFA}
m=\int dh \Psi_d(h) \tanh[\beta(qJm+h)],
\end{equation}  
where $q$ is the lattice coordination number, $J$ is the interaction constant
and $\beta=1/kT$ is the inverse temperature. It is 
well-known\cite{Pytte,Aharony} that for a single gaussian field distribution 
with dispersity $r=\delta$ the critical temperature $T_c$ decreases with 
increasing $\delta$, reaching $T_c=0$ for $\delta=\delta_c$ such that 
$qJ/\delta_c=\sqrt{\pi/2}$. For $\delta>\delta_c$ the system remains 
non-critical ($m=0$) at all temperatures. For an arbitrary field distribution $\Psi(h)$
the condition for such a zero $T_c$ point is given by \cite{Aharony}
\begin{equation}
\label{condition}
2qJ\Psi(h=0)=1
\end{equation} 
In the case of the sample averaged 
distribution $\Psi_d(h)$ (\ref{pd}) from the condition (\ref{condition}) we have found the following relation for the threshold width $a=a_c$ corresponding to the zero-temperature critical point  
\begin{equation}
\label{condition1}
a_c/\delta_c=c+(1-c)/(1+\Delta/a_c)=\frac{1+c\Delta/a_c}{1+\Delta/a_c}
\end{equation}
which implies $a_c<\delta_c$ suggesting that the zero-temperature 
critical point is shifted towards the region of a weaker field disorder. The 
phase diagram is given in Fig.~1 (the inset). It is seen that $T_c$
decreases with increasing distance $\Delta-a$ between the two sample 
groups. Even if we 
take a population with $a<<\delta_c$, such that all these samples
are critical (two phases are detectable), then the result seen on average 
depends on the population (magnitude of $c$ and $\Delta$) of non-critical 
($m=0$) samples. Therefore, even a small heterogeneity in the realizations
suppresses the tendency towards  the ferromagnetic ordering. 
 We would like to emphasize the  difference in the interpretation 
of this result from the  simulation/theoretical and the experimental points of 
view.  If the simulation samples are generated as distinct realizations, then 
testing each sample (e.g. by decreasing temperature and recording the 
magnetization) one would detect the criticality or not. Then the critical 
and non-critical samples could be considered as qualitatively different and 
taking an average over this group of dissimilar objects might seem to 
make little sense. Nevertheless, in order to mimic the 
behavior of a real self-averaging system (in the experimental sense) one has to 
perform such a realization averaging. On the other hand, if the realizations 
are thought to be macroscopic subsystems of a single sample, then they 
cannot be tested separately and the observed response is a superposition 
given by eqs.~(\ref{reals}), (\ref{obs}). In that case an experimentalist 
would conclude that the system is critical or not depending on the 
preparation details ($a,\Delta,c$). But, as we have already seen, the 
conditions for observing the criticality are quite strict. Small 
inhomogeneities in the realizations shift the transition to low temperatures, 
which might be difficult to reach.  
\subsection{Continuous sampling}
In practice it is difficult to expect that the realizations will fall
into a finite number of distinct groups. A more realistic sampling should 
involve a quasi-continuous sample distribution in some interval. In this 
context we consider a continuous step-wise distribution 
$\varphi(r)=\varphi_c(r)$ 
\begin{equation}
\label{sampling}
\varphi_c(r)=H(r)H(\Delta-r)/\Delta,
\end{equation}
where $H(r)$ is the Heviside step function. In this way we have
a set of samples with the gaussian field distribution $f(h|r)$ (\ref{fg}), 
whose dispersity is distributed between $r=0$ and $r=\Delta$ with equal 
probability. According to (\ref{reals}) the average over this set
is given by
\begin{equation}
\Psi_c(h)=\frac{1}{\sqrt{8\pi\Delta^2}}
E_1 \left(
\frac{h^2}{2\Delta^2} 
\right),
\end{equation}
where $E_1(x)$ is the exponential integral\cite{Abramowitz}. 

Let us first analyze the statistical features associated with different realizations. 
The distribution $\Psi_c(h)$ is displayed in Fig.~2 together with a gaussian 
distribution of the same width. It is seen that $\Psi_c(h)$ has a 
non-gaussian shape with a logarithmic singularity at $h=0$ and a 
well-pronounced tail favoring stronger fields $h$ (see the inset) with 
higher probability than it was in the case of the bimodal sampling. The 
striking difference in the sample and the sample-averaged statistics is 
worth of being emphasized. Namely, in each sample the field fluctuation is 
determined by a quenching path ${\overline{h^2}}-{\overline{h}}^2=r^2$ (the 
average is taken with $f(h|r)$, eq.~(\ref{fg})). The sample-averaged 
fluctuations (given by  $\Psi_c(h)$) are controlled by the sampling 
procedure ${\overline{h^2}}-{\overline{h}}^2=\Delta^2/3$. Moreover, the 
sample-averaged field fluctuations are coupled to the sample 
dispersity fluctuations 
$$
{\overline{h^2}}-{\overline{h}}^2=4
[{\overline{r^2}}-{\overline{r}}^2]=\Delta^2/3,
$$ 
where the correspondent averages are taken with the distributions
$\Psi_c(h)$ and $\varphi_c(r)$. Therefore, the sample averaging induces
specific fluctuations, which do not take place in separate samples. 
This reflects a superposition of different realizations contributing
to the overall result.

To see how does this affect the criticality we consider a uniform sampling distribution
with the gaussian width distributed between $r=a$ and $r=\Delta$, ($\Delta>a$).
\begin{equation}
\varphi(r)=\frac{1}{\Delta-a}H(r-a)H(\Delta-r)
\end{equation} 
In this way we avoid the counting of the "pure" states with $r=0$, for which $f(h|r=0)=\delta(h)$ and there is no field disorder.  
Then the realization averaged field distribution is given by
\begin{equation}
\Psi(h)=\frac{1}{\sqrt{8\pi(\Delta-a)^2}}
\left[
E_1 \left(
\frac{h^2}{2\Delta^2} 
\right)-
E_1 \left(
\frac{h^2}{2a^2} 
\right)
\right]
\end{equation}
From the condition (\ref{condition}) for the $T_c=0$ critical point we obtain
\begin{equation}
\label{condition2}
\frac{a_c}{\delta_c}=\frac{\ln{(\Delta/a_c)}}{\Delta/a_c-1}
\end{equation}
where, as before, $\delta_c$ is the threshold width of a single gaussian distribution.
Therefore, as in the case of the discrete sampling, $a_c/\delta_c\le 1$. This implies that
the lower border $a$ should be below the gaussian threshold $\delta_c$ at a "right" distance
determined by the maximal width $\Delta$. In other words, the transition is shifted towards
the region of weaker disorder. In order to study the phase behavior at non-zero temperatures 
we have again analyzed the MFA (\ref{MFA})
with $\Psi_d(h)$ being replaced by $\Psi(h)$. Our numerical analysis
reveals that the phase diagram is qualitatively similar to that obtained for the discrete
sampling (the inset in Fig.~1). Therefore, the ordered phase ($m\ne 0$) appears only if the
realization parameters $a$ and $\Delta$ are chosen in a proper way with respect to $\delta_c$.
Thus, changing the system composition with a small fraction of impurities may completely "eliminate" the ordered phase. 
 Note that quite recent 
experiments\cite{copolymers} on lamellar ordering transitions in doped block 
copolymers reveal that a very small concentration ($\approx 1\%$) of the 
doping agent suppresses the ordered phase.

One might argue that the 
quenched variable distributions used in the simulation or theoretical 
studies are already given as averaged over realizations $\Psi(h)$. Such that 
one works with a "representative" sample, trying to find general features, 
independent of the preparation details. This, however, rises a question on 
how to get it in practice (both in experiments and simulations). In other 
words, what should be the sample distributions $f(h|r)$ that, after the 
averaging with some $\varphi(r)$, leads to a given functional form for 
$\Psi(h)$. Let us assume that the sample average distribution 
$\Psi(h)=\Psi_g(h|\sigma)$ is gaussian with zero mean and dispersity 
$\sigma$, and that the sampling is again continuous - eq.~(\ref{sampling}). 
Then, following eq.~(\ref{reals}) we have to solve \begin{equation}
\Psi_g(h|\sigma)=\frac{1}{\sigma}\int\limits_{0}^{\sigma} dr f(h|r)
\end{equation}
with respect to $f(h|r)$. The solution is
\begin{equation}
f(h|r)=\left(\frac{h}{r}\right)^2 \Psi_g(h|\sigma=r).
\end{equation}
It is clear that in each sample, $r$, the field disorder is bimodal with
$f(h|r)$ having a minimum at $h=0$ and two maxima at $h_m=\pm r\sqrt{2}$.
It has been reported\cite{Aharony} that the bimodality (or more 
precisely, a minimum at $h=0$) is a prerequisite of a tri-critical behavior. 
If the maxima are separated enough (with increasing $h_m$), then in (at 
least) some fraction of the realizations a first-order transition is 
observable \cite{remark}.  Obviously, this effect is lost after the sample averaging. 
As is discussed above, with the gaussian 
$\Psi_g(h|\sigma)$ a second-order transition is observable depending on the
magnitude of $\sigma$.
A quite similar softening of a first-order transition has  been 
detected both experimentally\cite{expsoft} and theoretically
\cite{Soft}.  

It should be noted that the smearing of the singularities discussed 
here is not a peculiarity of the chosen model or probability distributions.
Let us assume that for a given realization $r$ the disorder-averaged 
quantity of interest $q(x|x_0)$ makes a step at $x=x_0=x_0(r)$. In the 
vicinity of the step we can approximate $q(x|x_0)=q_0H(x-x_0(r))$, where 
$H(x)$ is again the Heaviside step function. In different realizations (samples) 
the steps occur at different $x_0(r)$. This can be translated into a 
probability density $g(x_0|s)$ to find the realization with a given $x_0$, 
where $s$ is a parameter related to the sampling procedure. In particular, 
for the uniform sampling (\ref{sampling}) we can identify $s$ with $\Delta$. 
The realization averaging leads to 
\begin{equation} 
q(x|s)=\int\limits_{-\infty}^{+\infty} dx_0 g(x_0|s) 
q(x|x_0)=q_0\int\limits_{-\infty}^{x} dx_0 g(x_0|s) \end{equation} 
Since $g(x_0|s)$ is a probability density, then the last integral in the 
equation above is (by definition) a probability to find a given value of 
$x$. Therefore, unless $g(x_0|s)$ is singular, $q(x|s)$ varies smoothly in 
between $0$ and $q_0$. The same arguments apply also to divergent functions.
For instance, $dq(x|x_0)/dx=q_0\delta(x-x_0(r))$. Its realization average
\begin{equation}
\label{dq}
\frac{dq(x|s)}{dx}=q_0 g(x|s)
\end{equation} 
is not singular for any smooth realization (sample) distribution $g(x|s)$. 

\section{Conclusion}
We have found that the averaging over disorder realizations in the 
simulations or appearing in the context of the self-averaging 
hypothesis \cite{RF,SG} in experiments, is equivalent to 
crucial changes in the distribution of quenched variables. In this way a 
sample averaged distribution brings up new statistical features which are 
absent from the statistics of each sample. Consequently, the phase behavior 
seen on average could be remarkably different from that in a separate sample. 
It is found that the realizations need not be very different in order
to induce qualitative differences in the phase behavior.  
In particular, the sample averaging has a tendency 
to suppress both first- and second-order transitions, unless the ensemble preparation is done
under "right" conditions (e.g. eqs.~(\ref{condition1}), (\ref{condition2}) in the context of the present study). 
Based on this we
may argue that the absence of reliable evidences of a sharp phase 
behavior in the experiments\cite{copolymers,expsoft} with disordered systems 
can be explained by their self-averaging nature, involving a superposition of different disorder realizations. In this respect a development of   
 methods\cite{JPCB-const} which are not based on
the criticality could be quite promising for the characterization
of disordered media (e.g. porous or amorphous materials).

\newpage

\begin{figure}
\caption{Bimodal gaussian $\Psi_d(h)$ (solid) and 
 a gaussian (dashed) distributions of the same width.
 The inset displays the phase diagram for different distances
 between the gaussian populations, where $\delta_c$ is the threshold
 disorder width for a single gaussian (see text).}
\end{figure}
\begin{figure}
\caption{Continuous $\Psi_c(h)$ (solid) and 
 a gaussian (dashed) distributions of the same width.
 The inset displays $Log_{10}\times 10^{-2}$ of these distributions
 (the tails).}
\end{figure}

\begin{thebibliography}{99}
\bibitem{remark}
This has been shown \cite{Aharony} within a mean-field analysis which is
believed to be exact for the infinite range interactions. In the case of nearest neighor interactions a more refined analysis involving computer simulation techniques \cite{ceva,fytas}
does not give a definite answer.
\end{thebibliography}
\end{document}